\documentclass[a4paper,fleqn]{cas-dc}

\usepackage{listings}

\definecolor{codegreen}{rgb}{0,0.6,0}
\definecolor{codegray}{rgb}{0.5,0.5,0.5}
\definecolor{codepurple}{rgb}{0.58,0,0.82}
\definecolor{backcolour}{rgb}{0.95,0.95,0.92}

\usepackage[numbers]{natbib}

\def\tsc#1{\csdef{#1}{\textsc{\lowercase{#1}}\xspace}}
\tsc{WGM}
\tsc{QE}
\tsc{EP}
\tsc{PMS}
\tsc{BEC}
\tsc{DE}
\begin{document}
\lstdefinestyle{mystyle}{
    backgroundcolor=\color{backcolour},   
    commentstyle=\color{codegreen},
    keywordstyle=\color{magenta},
    numberstyle=\tiny\color{codegray},
    stringstyle=\color{codepurple},
    basicstyle=\ttfamily\footnotesize,
    breakatwhitespace=false,         
    breaklines=true,                 
    captionpos=b,                    
    keepspaces=true,                 
    numbers=left,                    
    numbersep=5pt,                  
    showspaces=false,                
    showstringspaces=false,
    showtabs=false,                  
    tabsize=2
}

\lstset{style=mystyle}

\let\WriteBookmarks\relax
\def\floatpagepagefraction{1}
\def\textpagefraction{.001}
\shorttitle{Porting WarpX to GPU platforms}
\shortauthors{A. Myers et~al.}

\title [mode = title]{Porting WarpX to GPU-accelerated platforms}                      
\author[1]{A. Myers}[orcid=0000-0001-8427-8330]
\author[1]{A. Almgren}[orcid=0000-0003-2103-312X]
\author[1]{L. D. Amorim}[orcid=0000-0002-1445-0032]
\author[1]{J. Bell}
\author[5]{L. Fedeli}[orcid=0000-0002-7215-4178]
\author[2,1]{L. Ge}
\author[1]{K. Gott}[orcid=0000-0003-3244-5525]
\author[3]{D. P. Grote}[orcid=0000-0002-4057-8582]
\author[2]{M. Hogan}
\author[1]{A. Huebl}[orcid=0000-0003-1943-7141]
\author[1]{R. Jambunathan}[orcid=0000-0001-9432-2091]
\author[1]{R. Lehe}[orcid=0000-0002-3656-9659]
\author[2]{C. Ng}
\author[1]{M. Rowan}[orcid=0000-0003-2406-1273]
\author[1]{O. Shapoval}[orcid=0000-0003-4003-4507]
\author[4]{M. Th\'evenet}[orcid=0000-0001-7216-2277]
\author[1]{J.-L. Vay}[orcid=0000-0002-0040-799X]
\author[5]{H. Vincenti}[orcid=0000-0002-9839-2692]
\author[1]{E. Yang}[orcid=0000-0002-9319-4216]
\author[5]{N. Za\"{i}m}[orcid=0000-0002-9582-5894]
\author[1]{W. Zhang}[orcid=0000-0001-8092-1974]
\author[1]{Y. Zhao}[orcid=0000-0003-4362-3630]
\author[1]{E. Zoni}[orcid=0000-0001-5662-4646]

\address[1]{Lawrence Berkeley National Laboratory, Berkeley, CA 94720, USA}
\address[2]{SLAC National Accelerator Laboratory Menlo Park, CA 94025,
USA}
\address[3]{Lawrence Livermore National Laboratory, Livermore, CA 94550, USA}
\address[4]{Deutsches Elektronen Synchrotron (DESY), Hamburg, Hamburg 22607, Germany}
\address[5]{LIDYL, CEA-Universit\'e Paris-Saclay, CEA Saclay, 91 191 Gif-sur-Yvette, France}

\begin{abstract}
WarpX is a general purpose electromagnetic particle-in-cell code that was originally designed to run on many-core CPU architectures. We describe the strategy, based on the AMReX library, followed to allow WarpX to use the GPU-accelerated nodes on OLCF's Summit supercomputer, a strategy we believe will extend to the upcoming machines Frontier and Aurora. We summarize the challenges encountered, lessons learned, and give current performance results on a series of relevant benchmark problems.
\end{abstract}

\begin{keywords}
Exascale Computing \sep Particle-in-Cell Methods \sep Accelerator Modelling
\end{keywords}

\maketitle

\section{Introduction}

WarpX \cite{WarpX} is a fully electromagnetic Particle-in-Cell (PIC) code that is being developed as part of the US Department of Energy's Exascale Computing Project \cite{ECP}. Originally designed for particle accelerator modelling, and in particular for the study of laser- and beam-driven wakefield accelerators, it has also been used to study several other topics in the field of laser-plasma interaction, such as probing the onset of Quantum Electrodynamics (QED) in extreme fields and laser-ion acceleration.

WarpX implements the well-known electromagnetic PIC method for solving the motion of relativistic, charged particles in the presence electromagnetic fields. In addition, it also includes support for many advanced features, such as: perfectly matched layers (PMLs) \cite{Shapoval2019}, a pseudo-spectral (PSATD) Maxwell solver \cite{VincentiCPC2016}, and multi-physics options such a ionization, QED pair creation \cite{Nikishov1970,Gonoskov2015}, a QED vacuum polarization solver \cite{carneiro2017quantum}, and binary collisions \cite{Perez2012}.
WarpX currently supports 2D, 3D and azimuthally decomposed geometries \cite{Lifschitz2009} and has the ability to operate in a Lorentz-boosted reference frame \cite{Vayprl07}. It also includes support for mesh refinement and dynamic load balancing through the AMReX library \cite{AMReX, zhang2020amrex}.

WarpX was originally designed with many-core architectures in mind. While the high-level operations such as time-stepping and MPI parallelization were implemented in C++ using AMReX data structures, the core PIC operations, such as current deposition, field gathering, and various particle pushers and field solvers, were handled by the PICSAR library of Fortran kernels \cite{VincentiSIMD}. These routines were highly optimized for the Intel's Knight's Landing (KNL) architecture found on supercomputers such as NERSC's Cori and ALCF's Theta platforms and featured hand-vectorized versions of the core PIC operations. Related works presenting early adoptions of accelerator hardware in PIC codes are presented in Refs. \cite{Burau2010,Bussmann2013,Surmin2016,VincentiSIMD}.

As the ECP focus shifted towards GPU-based machines such as Summit, Frontier, and Aurora, the question naturally arose about what to do with the Fortran kernels in PICSAR. CUDA Fortran provided a way forward for platforms with NVIDIA hardware such as Summit \cite{CORAL}, but it was not clear what support for Fortran would look like on AMD or Intel hardware. Likewise, OpenACC provided an easy-to-use model for offloading Fortran routines to NVIDIA GPUs, but again it was not clear how that would work with non-NVIDIA GPUs and compiler support was limited to a single vendor. OpenMP offered better prospects for portability; however, early implementations suffered performance problems when compared to OpenACC on NVIDIA hardware.

Ultimately, the choice was made to port the PIC kernels in PICSAR from Fortran to C++, and to offload kernels using either CUDA, HIP, or DPC++, depending on whether NVIDIA,  AMD, Intel hardware is targeted. This removed any need for mixed language programming, which adds substantial complication to the codebase and also defeats important compiler optimizations such as inlining. Additionally, C++ usually gets better and, importantly, earlier support from vendors, owing to its prominence in industry relative to Fortran. Finally, CUDA, HIP, and DPC++ offer a relatively consistent programming model across all three target platforms. Any implementation differences between the three could be entirely hidden in a performance portability layer; in our case, in the \verb+ParallelFor+ routines in AMReX (see Section \ref{sec:ParallelFor}). In this manner, NIVIDA, AMD, and Intel GPUs could all be supported, with little to no change to the WarpX code required.

Today, WarpX is a C++14 application with an optional, standardized Python interface (PICMI) that can be used to drive simulations interactively\cite{PICMI}. It runs on NVIDIA, AMD, and Intel GPUs, in addition to many-core CPU architectures such as KNL and Fujitsu's A64FX processors. The core of the GPU support is the \verb+ParallelFor+ kernel launching method from AMReX. In what follows, we give a brief overview of the features in AMReX that WarpX uses to enable parallelization and GPU support.  Then, we summarize several key lessons learned from the experience of scaling up the code on Summit. In particular, we discuss the importance of memory management, the importance of optimizing for memory footprint as well as run time, the importance of optimizing parallel communication routines, and the importance of properly utilizing the memory hierarchy. Finally, we will present weak and strong scaling results from a uniform plasma problem setup and performance numbers obtained on a plasma acceleration stage benchmark problem.

\section{Parallelization}

WarpX leverages AMReX for parallelization and GPU offloading. AMReX is a framework for building block structured, adaptive mesh refinement (AMR) applications that is being developed as part of the Exascale Computing Project. It provides distributed data containers for storing and iterating over mesh and particle data defined on AMR hierarchies, tools for performing parallel communication, inter-level operations, linear solvers, and support for embedded boundaries via a cut cell approach. In what follows we summarize the main features of the hierarchical parallelism model from AMReX that WarpX uses to run on multi-node CPU and GPU platforms.

\subsection{Domain Decomposition} 

WarpX makes use of the AMReX-provided tools for describing block-structured AMR hierarchies. From the simplest to the most complex, these are the \texttt{IntVect}, which describes a point in an integer index space; the Box, which describes a region in the same index space and consists of low- and high- end \texttt{IntVect}s plus a third \texttt{IntVect} that describes the staggering (i.e. is the box cell-, node- face, or edge-centered); the \texttt{BoxArray}, which describes a collection of Boxes at a given level of refinement; and the \texttt{Distribution\-Mapping}, which describes how each of those Boxes is mapped to MPI ranks. \texttt{Vector<BoxArray>} and \texttt{Vector<Distribution\-Mapping>} then describe the mesh hierarchy across multiple levels of refinement. When used to describe a block in an AMR hierarchy like this, we use the term `grid` interchangeably with \texttt{Box}. 
	
\texttt{DistributionMapping}s can be user-generated, or AMReX can generate them for a given \texttt{BoxArray} using a number of algorithms: round robin, knapsack, and space filling curve. By default, the boxes in WarpX are assigned to MPI ranks according to the space filling curve algorithm, which attempts to put nearby boxes on the same rank. When dynamic load balancing is employed, the user can select to use either space filling curve, which attempts to maintain spatial locality, or knapsack, which provides the most flexibility to achieve a balanced work distribution.

\subsection{Mesh and particle data structures}
	
The basic mesh data structure in AMReX is the \texttt{FArrayBox}, which is a multi-dimensional array of floating point values defined on a given Box. \texttt{FArrayBox}es can be single- or multi-component and used to represent scalar or vector physical quantities. In the multi-component case, the components are stored in a Struct-of-Array style. A distributed collection of \texttt{FArrayBox}es defined on a given \texttt{BoxArray} and \texttt{Distribution\-Mapping} is called a \texttt{MultiFab}. The \texttt{MultiFab} data structure is what stores the core mesh data fields in WarpX - the electric and magnetic fields, the current density, and so forth.

The core particle data structure in AMReX is the \texttt{Particle}, which consists of a collection of real and integer components. In WarpX, the majority of these components are stored in Struct-of-Array style. The exceptions are the particle positions and a 64-bit integer identification number, which are stored together in a separate \texttt{struct}. 
	
A \texttt{ParticleContainer} is a distributed collection of particles associated with a given AMR hierarchy. In WarpX, each particle species (driver beam, plasma elections, ions, etc...) is stored in a separate \texttt{ParticleContainer}. Particles are assigned to AMR levels and grids based on their physical positions. The particles in the \texttt{ParticleContainer} can then be looped over grid-by-grid, and PIC operations such as field gathering, particle pushing, and current deposition can be performed.

\subsection{Hierarchical Parallelism}
The above data structures naturally lend themselves to an ``MPI+X" hierarchical approach to parallelism, where "X" is one of OpenMP, CUDA, HIP, or DPC++ for on-node accelerated compute. Boxes are assigned to MPI tasks, and we typically use a form of over decomposition so that each MPI tasks is responsible for processing multiple boxes. This allows for more flexible grid structures and also aids load balancing via swapping boxes across ranks. AMReX provides \texttt{MultiFab} and \texttt{Particle\-Container} iterator objects that can instruct each rank to loop over their local grids, processing each one in turn. When processing individual grids, an accelerated compute backend such as OpenMP or CUDA can be selected to perform the actual computations. See the section on \nameref{sec:ParallelFor} for more information.

\subsection{Parallel Communication routines}
	
Communicating mesh and particle data between MPI ranks is handled by the AMReX framework. In particular, WarpX makes use of the following parallel communication routines:
	
\begin{itemize}
\item \textbf{FillBoundary}: This method is used to fill guard cells for the mesh data (e.g. the electric and magnetic field components, and current / charge densities). It fills the data in the guard cells with the (possibly more recent) data from the corresponding valid cells. Here, "valid" refers to cells that are uniquely owned by the grid in question, as opposed to ghosted copies that may exist on other grids / MPI processes.
	
\item \textbf{SumBoundary}: This operation is analogous to \texttt{Fill\-Boundary}, except that instead of copying from valid to guard, it takes the values in the ghost cells and adds them to corresponding valid cells. This is useful when doing current and charge deposition operations on particles that are near the edge of grid boundaries. These particles add some of their weights to guard cells, and these contributions are summed to the proper valid cell by the \texttt{SumBoundary} method.
	
\item \textbf{SyncNodal}: This function is used when staggered or node-centered grids are employed to represent physical quantities. For example, when using the standard Finite-Difference Time-Domain (FDTD) Maxwell solver, WarpX uses the Yee \cite{Yee} discretization of the field variables, which stores the magnetic field components on cell faces, and the electric field and current density components at cell edges. In this case, some points in the discretization are represented on multiple grids and potentially multiple MPI ranks. Note that, unlike in the cell-centered case, no one grid can be said to u\-nique\-ly ``own" these points. To prevent spurious numerical effects, it is necessary to synchronize these shared nodal points so that they have exactly the same values to machine precision. The \texttt{SyncNodal} method accomplishes this. Several different options for deciding which value to use are implemented, e.g. simple or weighted averaging. By default, WarpX simply chooses an arbitrary value as the `winner` and overrides the others.

\item \textbf{ParallelCopy}: This is the most general form of parallel communication for mesh data in AMReX. It performs copy on intersection from one \texttt{MultiFab} to another, even when those \texttt{MultiFab}s have different \texttt{Box\-Array}s and \texttt{DistributionMapping}s. This is needed when, for example, copying data between different levels of refinement, performing regridding or load balancing operations, and when copying data between the PML grids and the rest of the domain.
	
\item \textbf{Particle Redistribution}: This refers to putting particles back on the proper level and grid after they have been pushed. AMReX includes two versions of this operation, one in which the particles are assumed to only move between neighboring ranks, and another in which they are allowed to move between any two ranks in the MPI communicator. The former, local version is the one used most often during normal time stepping, while the latter version is used when performing load balancing.
\end{itemize}
	
These parallel communication routines have been optimized for hybrid CPU/GPU platforms, in particular, Summit. All run fully on the GPUs, meaning that they do not trigger any unnecessary host/device copies of mesh or particle data. Communication has been refactored to reduce the number of GPU kernels launched, in particular when performing packing and unpacking operations on MPI send and receive buffers (see Section \ref{sec:comm_opt}). Finally, WarpX can take advantage of gpu-aware MPI implementations that can operate directly on device data pointers, if one is available. This operation can be enabled using a runtime option. When turned off, right before the MPI sends are performed, the data to be sent is copied into pinned memory buffers on the host and the MPI exchanges are made between host memory buffers instead.
	
\subsection{ParallelFor}
\label{sec:ParallelFor}

The core of the GPU support in WarpX consists of a series of \texttt{ParallelFor} functions provided by the AMReX framework. These are similar to those provided by the performance portability layers Kokkos \citep{kokkos} and RAJA \citep{raja}, but have been tailored towards the needs of structured grid applications. These functions separate the details of how the loop is performed from the loop body, which is supplied as a C++ lambda function describing the operation done on each element. The \texttt{amrex::ParallelFor} loops are translated into either GPU kernel launches or normal host \texttt{for} depending on a compile-time option. Using this approach, a single code base can be maintained that can run on CPUs and on multiple GPU platforms. Note, however, that \texttt{amrex::ParallelFor} does not include OpenMP parallelization - that is included in the outer, \text{MFIter} level instead (see Listing \ref{lst:mfiter_parfor}).

Listings \ref{lst:parallelfor_box} and \ref{lst:parallelfor_particle} show two examples of \texttt{ParallelFor}. The first specifies the loop bounds using an AMReX \texttt{Box} object, which results in a 3-dimensional loop over the cells in the box. The second shows a one-dimensional \texttt{ParallelFor}, which loops over the particles in a grid. 

\begin{lstlisting}[language=C++,frame=single,basicstyle=\footnotesize,label={lst:parallelfor_box}, caption={A \texttt{ParallelFor} routine operating on a single box of mesh data. In this example case, the threading will be performed over the cells of a 3-dimensional box. AMReX arrays use Fortran index order.}]
amrex::ParallelFor(bx,
    [=] AMREX_GPU_DEVICE (int i, int j, int k)
    {
        dstarr(i,j,k,0) = srcarr(i,j,k,0);
    });
\end{lstlisting}

\begin{lstlisting}[language=C++,frame=single,basicstyle=\footnotesize,label={lst:parallelfor_particle}, caption={A one-dimensional \texttt{ParallelFor} used to thread over all the particles in a grid.}]
amrex::ParallelFor(np,
    [=] AMREX_GPU_DEVICE (int i) 
    {
        amplitude[i] = 0.0_rt;
    });
\end{lstlisting}

\subsection{Reductions}
\label{sec:Reductions}

Parallel reductions are used in several places in WarpX; for example, in diagnostic functions for particles and beams that act as an \textit{in situ} data reduction technique. AMReX provides functions implementing these reduction operations including both ``off-node" reductions over MPI ranks and ``on-node" reductions using either CPU threads or on-device acceleration. On GPUs, these functions make use of vendor-supported libraries such as CUB \cite{CUB} and rocPRIM \cite{rocPRIM} when possible, and otherwise fall back on AMReX's own implementation based on on warp-level primitives such as CUDA's \texttt{\_\_shfl\_down} operation and atomics.

A feature of the AMReX implementation of parallel reductions is that it provides an API for performing multiple reductions in one pass on any combination of data types and reduction operators. When running on GPUs, all these operations would be done in a single kernel launch. These reductions operations have been tested and implemented for CUDA, HIP, and DPC++, as well as on CPU platforms using OpenMP. 

\section{Lessons from Summit}
\label{sec:lessons}
In the following section, we summarize some key lessons learned from our experience of porting WarpX to Summit. These fall under three main areas: issues relating to memory management and overall footprint, issues relating to parallel communication, and finally, the importance of cache utilization on GPU platforms.

\subsection{Memory Optimization}
\label{sec:mem_opt}
With the trend towards GPU computing, the importance of optimizing codes for memory consumption has increased. Consider the example of Summit. Summit has 4608 nodes, each of which has 608 GB of host memory (512 DDR4 + 96 HBM2), for a system total of 2.8 Petabytes. This is considerably more than Cori's KNL nodes, which have a total aggregate memory of 1.1 Petabytes. However, if we consider only device memory, each Summit node has 6 NVIDIA V100 GPUs with 16 GB of memory each, for a total of only 440 TB, substantially less than Cori phase II. This means that, provided that one wants to run in a mode in which your problem entirely fits on the GPUs (which is desirable considering the performance penalties associated with frequent host / device data transfers), one actually cannot run as big of a problem on Summit than one could fit on Cori. This makes reducing the memory footprint of a simulation code quite impactful in terms of enabling production-level science calculations. 

Reducing the memory footprint can have performance implications as well. Originally in WarpX, every particle stored persistent values for the electric and magnetic fields interpolated to the particle's position. In addition to the storage overhead, these values need to be communicated every time particles change MPI domains, and shuffled around in memory every time particles are sorted (see Section \ref{particle_sorting}). Additionally, if the performance of a GPU kernel is memory-bound, meaning that its performance is limited by the rate at which data can be transferred from main memory to the streaming multiprocessors on the GPUs, then increasing the arithmetic intensity of those kernels by streaming less data and recomputing values on-the-fly can improve their overall performance. 
	
Recently, WarpX removed the persistent electric and magnetic fields at the particle positions in favor of re-gathering these values inside GPU kernels as they are needed. For this, the field gathering and particle pushing kernels were fused together in the PIC loop, resulting in less data that needed to be streamed to the processors in a given timestep. In addition to reducing the memory footprint by a factor of $\approx1.6$, this also led to a $\approx25$\% percent speedup in the overall runtime on several key benchmarks. When the field values at the particle positions are needed more than once in a step, as, for example, when modelling additional effects such as ionization or using certain diagnostics, the gather operation is simply performed multiple times. This kind of optimization has the added benefit that it will likely improve CPU performance as well for computational kernels that are memory bound, which includes most of the operations performed in WarpX on modern CPU hardware.

Finally, we are currently exploring other means of reducing the overall memory footprint of WarpX, including exploiting single / mixed precision and employing compression.

\subsection{Memory Arenas}
\label{sec:mem_arena}
Dynamic memory allocation is many times more expensive on GPU than CPU architectures. This fact, combined with common programming patterns involving temporary variables, can lead to drastic performance penalties on GPU systems. For example, consider the code in Listing \ref{lst:mfiter_parfor}. This snippet demonstrates how to loop over mesh data using the AMReX data structures. The \texttt{MFIter} object instructs each MPI rank to loop over the grids it owns. For each grid, we resize a temporary scratch space called \texttt{tmp}, then launch a \texttt{ParallelFor} kernel to do some calculations using it. The \texttt{Elixir} is not essential to the point, but it keeps the scratch space alive in memory until the kernel is finished working with it - this is needed due to the asynchronous nature of GPU kernel launches. If every call to \texttt{resize} the buffer ended up triggering \texttt{cudaMalloc} and \texttt{cudaFree} calls, this could easily end up becoming the dominant cost of this routine. Another place this comes up is in out-of-place sorting and partitioning operations, which require a temporary buffer in which to store the result.

One way to mitigate this is to refactor application codes to keep temporary buffers alive in memory instead of letting them go out of scope. However, this is error-prone and labor-intensive. Instead, AMReX provides a number of memory pool classes termed ``\texttt{Arena}s", which allocate memory in large chunks and dole out pieces of it as the application runs. Thus, even though WarpX makes frequent use of temporary variables, during most time steps that are no calls to \texttt{cudaMalloc} or \texttt{cudaFree}. 

These \texttt{Arena}s have a number of different options for managing memory fragmentation; currently, the default in AMReX is to use a "first fit" strategy. AMReX provides memory arenas that use host, device, pinned, and managed memory. WarpX uses these Arenas for all of its mesh and particle data structures. 

By default, when running on NVIDIA GPUs, WarpX places all of its core data in an Arena that uses managed memory. However, this feature is optional, and we have tested the performance implications of this compared to placing the data in the device Arena on the 1-node version of the test problem presented in Section \ref{sec:weak}. Overall, the overhead associated with using unified memory appears to be less than $0.2\%$ on NVIDIA V100, likely because WarpX initializes all data on the device and AMReX uses \texttt{cudaMemAdvise} to set the preferred location of the managed memory \texttt{Arena} to ``device."

In addition to the managed and device memory arenas, WarpX also uses a pinned memory arena in a few places where frequent device-to-host transfers are needed: first, when performing MPI communication when GPU-aware is not available or is switched off (see Section \ref{sec:comm_opt}), and second, when copying simulation data to host for IO. 

\begin{lstlisting}[language=C++,frame=single,basicstyle=\footnotesize,label={lst:mfiter_parfor},caption={Example of an \texttt{MFIter} loop with a \texttt{ParallelFor} inside. This code can be compiled to run on CPUs with OpenMP or GPUs with CUDA, HIP, or DPC++.}]
#ifdef AMREX_USE_OMP
#pragma omp parallel if (notInLaunchRegion())
#endif
{
    FArrayBox tmp;
    for (MFIter mfi(mf, TilingIfNotGPU()); 
        mfi.isValid(); ++mfi)
    {
	    const Box& bx = mf.tilebox();
	    tmp.resize(bx);
	    Elixir eli = tmp.elixir();
	    auto const& tmp_arr = tmp.array();

	    amrex::ParallelFor(bx,
	    [=] AMREX_GPU_DEVICE (int i, int j, int k)
        {
	        compute_tmp(i, j, k, tmp_arr);
        }
    }
}
\end{lstlisting}

\subsection{Communication Optimization}
\label{sec:comm_opt}
Once the initial port of WarpX to NVIDIA GPUs was complete, the initial experience was that compute kernels such as current deposition and field gathering were much faster on V100 hardware than on KNL. However, the same was not true for the AMReX parallel communication routines. The primary reason for this was that the parallel communication routines involved many small, "copy on intersection" routines between neighboring boxes, especially when packing and unpacking MPI send and receive buffers. These operations involved little to no computation but launched many small kernels that packed and unpacked data buffers. Thus, the dominant cost in these routines was the latency associated with the kernel launches, which could be fused into a fewer number. After optimization, each MPI rank makes only 1 kernel launch to pack and unpack its MPI buffers, which led to greatly improved performance on Summit.

Additionally, the parallel communication routines in AMReX were restructured to take advantage of GPU-aware MPI implementations, in which pointers to device memory can be passed directly to MPI function calls without any need to explicitly transfer data to the host first. This feature is optional; in the event that it is switched off, data in MPI buffers is copied to pinned memory before it is passes to MPI. Note that to date, we have not seen large performance benefit from using GPUDirect on Summit.

\subsection{Cache utilization}

As with CPU-based many-core architectures, rearranging computations so that they properly exploit the memory hierarchy can lead to significant performance increases on V100 GPUs. In this section, we discuss a case-study in this effect - specifically, on how periodic sorting of particle data, so that it is processed in a cache-friendly way, can greatly improve the performance of PIC operations like field gathering and current deposition on V100. First, however, we will describe the current deposition algorithm we use, and how it differs between CPU and GPU runs, in more detail. 

\subsubsection{Current Deposition}

In PIC codes, most operations are straightforward to parallelize, since particles can be threaded over and processed independently without needing to worry about potential race conditions. Charge and current deposition operations, however, require special consideration, since when threading over particles there is the potential for collisions as multiple threads may attempt to update the same cell simultaneously. Note that, while most work loops in WarpX are performance portable between CPU and GPU architectures in the manner of Listing \ref{lst:mfiter_parfor}, the current deposition step is the one area where it does maintain two different code paths through the use of \texttt{\#ifdef} preprocessor directives. This is because the current deposition is a particularly important routine in terms of computational cost, and because fundamentally different approaches need to be selected depending on whether we are running with OpenMP or CUDA/HIP/DPC++ as the parallel backend, for reasons discussed below.

In WarpX, our approach to concurrent scatter operations in particle deposition kernels varies  With OpenMP, the particles on a grid are sorted onto smaller sub-regions called tiles. This sorting achieves two things: first, it is good for data locality, because particles in the same tile are likely to contribute to cells that are near each other in memory, and second, because it allows OpenMP threads to be mapped to tiles, which can then be processed simultaneously. Each OpenMP thread deposits particles onto its own, private deposition buffer with enough cells to capture the support of all the particles on the tile. There is no need for atomics at this stage, since each thread has its own buffer. After deposition onto the buffer is complete, the buffer values are atomically added to the values for the full grid using atomic writes. Thus atomics are only needed on a per-cell basis, not a per-particle basis. 

The above algorithm requires having few enough OpenMP threads that each one can have its own private deposition buffer large enough to cover the support of all the particles in a tile, a strategy sometimes called \emph{data duplication}. While this is possible when running on a few (say, 8-32) OpenMP threads, it becomes infeasible on GPUs, which can easily have tens of thousands of concurrently active threads. Thus, when running on GPUs, we dispense with thread-private buffers and perform atomic writes directly to global memory for each particle. The performance of global atomics on V100 is quite good; however it still beneficial to perform some sort of sub-box level sorting for the particles on a grid to achieve better data locality. Unlike with the OpenMP algorithm, this sorting process does not need to be done every time step, since we only need an approximate ordering to significantly improve cache re-use, not a strong guarantee that all the particles being processed by an OpenMP thread are inside the thread's buffer region. WarpX has experimented with thread-block level buffers in GPU shared memory to reduce the number of global atomic updates; however, none of these methods have been faster than the one described above. This details of particle sorting on GPUs are discussed in the next section.

\subsubsection{Particle Sorting}
\label{particle_sorting}

\begin{figure}[h]
\caption{\label{bin_size} The effect of sorting interval (i.e., sorting every $N$ time steps) and sort bin size on the overall performance on a uniform plasma benchmark. The $x$-axis shows the sort interval, while the $y$-axis shows the overall time to take 100 steps, including the cost of the sorting. A sort interval $>100$ means that the particles are never re-sorted during the run.}
\centering
\includegraphics[width=0.5\textwidth]{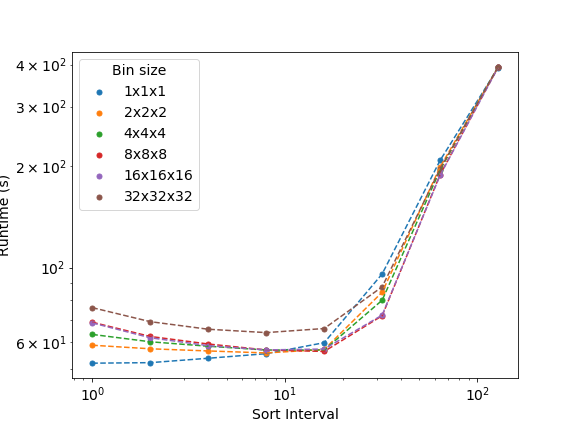}
\end{figure}

\begin{figure}[h]
\caption{\label{deposition_roofline} Roofline analysis of the 3rd-order Esirkepov current deposition \cite{Esirkepov2001} kernel in WarpX on a single V100 GPU, with and without particle sorting. In the memory streaming limit, three different lines are shown, corresponding to the bandwidths of the L1 and L2 caches as well as that for the main high-bandwidth memory (HBM) on the GPU. Likewise, in the compute-bound regime, two different values are used for the peak floating point performance: both with and without taking advantage of fused multiply-add instructions. The arithmetic intensity (A.I.) is measured three times for each kernel, using the memory traffic for each level of the memory hierarchy. For the sorted version, the fact the A.I. is significantly lower for the L1 and L2 data points shows that we are getting substantial reuse in both levels of cache. Conversely, the fact that the data points are all on top of each for the unsorted run indicates that without sorting, the degree of reuse is poor.}
\centering
\includegraphics[width=0.5\textwidth]{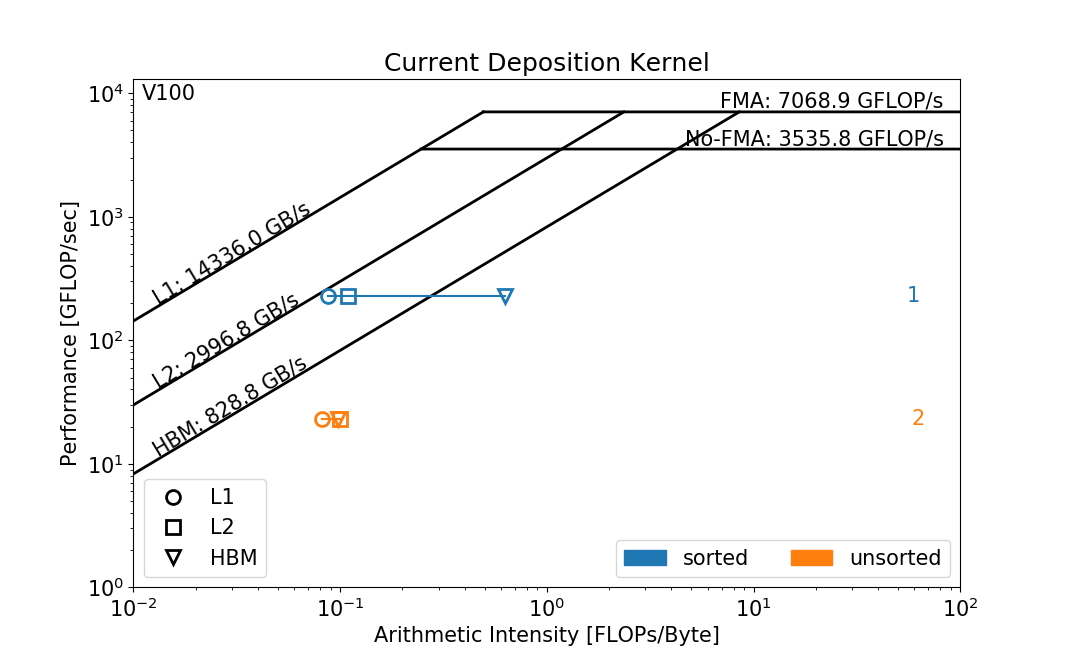}
\end{figure}

\begin{figure}[h]
\caption{\label{gather_roofline} Same as Figure \ref{deposition_roofline}, but for the fused gather and push kernel in WarpX. Again, there is substantial cache reuse when sorting is employed, although for this kernel performance still appears to be limited by HBM bandwidth, even with sorting.}
\centering
\includegraphics[width=0.5\textwidth]{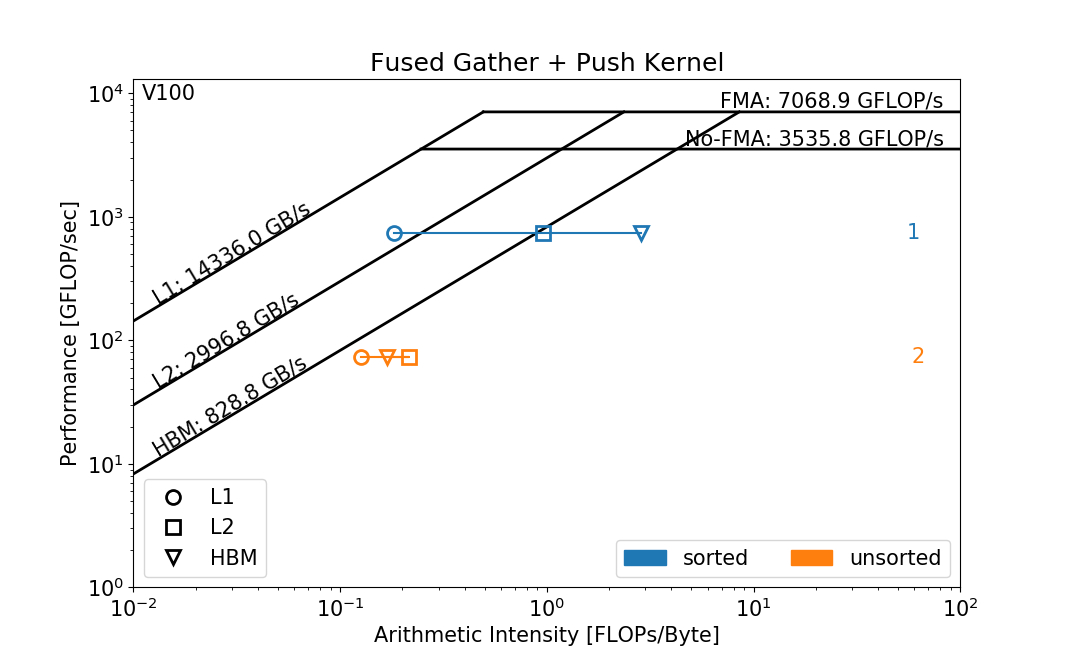}
\end{figure}

Periodic sorting of the particles on each grid by their spatial locations so that particles that are close to each in memory also interact with cells that are close to each in memory exploits the memory hierarchy on the GPUs more effectively than processing them in an unordered fashion. This is particularly true in the case that particles are moving with high velocities, such that they frequently change cells. In that case, even if particles are sorted at a particular time, they will rapidly become disordered, leading to significant performance degradation in the particle-mesh operations.
	
Note that we differentiate between binning, which computes a permutation array that assigns particle indices to cells with user-defined bin size, and sorting, which uses this permutation array to actually reorder the particle data in memory. Cache utilization requires full sorting, but for many operations simply knowing the cell-sorted indices is sufficient. AMReX provides a GPU-capable implementation of the counting sort operation that can be used to perform both of these operations. Internally, it is built using a GPU implementation of parallel prefix sum, which is based on Ref.\,\cite{Merrill2016SinglepassPP} and works on NVIDIA, AMD, and Intel GPUs.

In addition to the presented cache-utilization optimization, sorting and/or binning particles is needed for the modeling of particle-particle interactions. The PIC method by default only models particle-mesh interaction and mesh updates. WarpX implements binary collisions, which depend on a prior binning of neighboring particles, to address various applications in accelerator and beam physics.

Figure \ref{bin_size} shows the results of a parameter study in which the bin size and sorting interval were varied. For example, a bin size of $2x2x2$ and sorting interval of $4$ means that particles were sorted into $2x2x2$ supercells every 4 timesteps. On this problem, the optimal sorting is to sort by cell (i.e. a bin size of $1x1x1$ every time step, and the difference between sorting optimally and not sorting at all is a factor of $\approx 7.5$, with most of the improvement comings from the current deposition and fused gather and push kernels. However, the very frequent sorting interval for this problem is a special, because the particles in this problem change cell more often than in most WarpX applications. Currently, the default in WarpX, used throughout Section \ref{sec:performance}, is to sort the particles by their PIC cell every 4 time steps. 

Note that, although the \texttt{Redistribute()} function in AMReX does not maintain this cell-sorted order for particles that left one grid and been migrated to another, this only applies to particles that have changed grids - typically only a small subset of the total that are near the ``surface". The bulk of the particles on a grid will maintain their sorted order in between \texttt{Redistribute()} calls.

Figures \ref{deposition_roofline} and \ref{gather_roofline} show the results of a roofline analysis \citep{Sam} on the current deposition and fused gather and push kernels in WarpX, which are the two most computationally expensive operations. Our analysis followed the methodology of \cite{Yang}. For this test, we used a uniform plasma setup with 8 particles per cell and gave the particles a large thermal velocity, so that they frequently change cells. To rule out any transient effects, we ran the problem for a total of 100 steps and only profiled the last one. 

The roofline analysis reveals three things. First, as already demonstrated, sorting the particles gives significantly better performance on V100 GPUs than not sorting them. Second, the fact that the arithmetic intensity measured using the memory bandwidth for the L1 and L2 caches is significantly lower than for HBM indicates that, in the sorted run, we are getting significant reuse in both of these levels of cache. Third, the arithmetic intensity for the current deposition for the sorted run is right up against the streaming limit for the L2 cache. This indicates that the performance of this kernel is now limited by the L2 cache bandwidth. Gather and push, on the other hand, is likely still limited by HBM bandwidth. Taken together, these results suggest that these kernels should get significantly better performance on the A100, which has a larger L2 cache and higher HBM bandwidth than the V100.

Finally, we note that some PIC codes, such as PIConGPU \cite{Bussmann2013}, achieve a similar effect by explicitly using shared memory to cache the electric and magnetic fields for nearby particles during field gathering, and by using it as a write buffer when performing current deposition. We have experimented with this approach and have thus far not seen an advantage to doing so. However, work on this front is ongoing. We note that our approach achieves a similar caching effect by implicitly relying on L2 rather than explicitly managing the contents of shared memory buffers.

\section{Performance Results}
\label{sec:performance}
In this section, we give current performance results on Summit for two key benchmark problems. We concentrate on two areas - the scaling of the code on a uniform plasma test case and the performance on a plasma accelerator benchmark problem.

\subsection{Uniform Plasma Scaling}
\subsubsection{Weak scaling study}
\label{sec:weak}
In order to test the scaling of WarpX in an idealized setting, as well as to gauge the speedup associated with using accelerated nodes, we have performed a weak scaling study using a uniform plasma setup on OLCF's Summit supercomputer. The base case for this scaling study used a 256 x 256 x 384 domain with a box size of 128$^3$ and ran on 1 Summit node; thus, on the GPU-accelerated runs, each GPU was responsible for processing two 128$^3$-sized boxes. Particles were initially distributed uniformly with 8 particles per cell. We used the standard Yee FDTD solver for these runs, with Esirkepov current deposition and third order shape functions. For the weak-scaling study, the number of Summit nodes were doubled with the number of cells (and particles therein) in the x-, y-, or z- directions, while holding everything else constant, maintaining a constant workload per node. We continued this process up to 2048 nodes - about half of the Summit machine. Overhead associated with time spent in problem initialization, memory allocation, etc., was minimized by running for a total of 100 steps.

\begin{figure}[h]
\caption{\label{weak_scaling} Results of a weak scaling study on a uniform plasma setup on Summit. The x-axis shows the number of Summit nodes, while the y-axis is the number of particles advances per nanosecond. Both the CPU and GPU versions of the code scale well, and the overall speedup associated with using the accelerators is $\sim 30$.}
\centering
\includegraphics[width=0.45\textwidth]{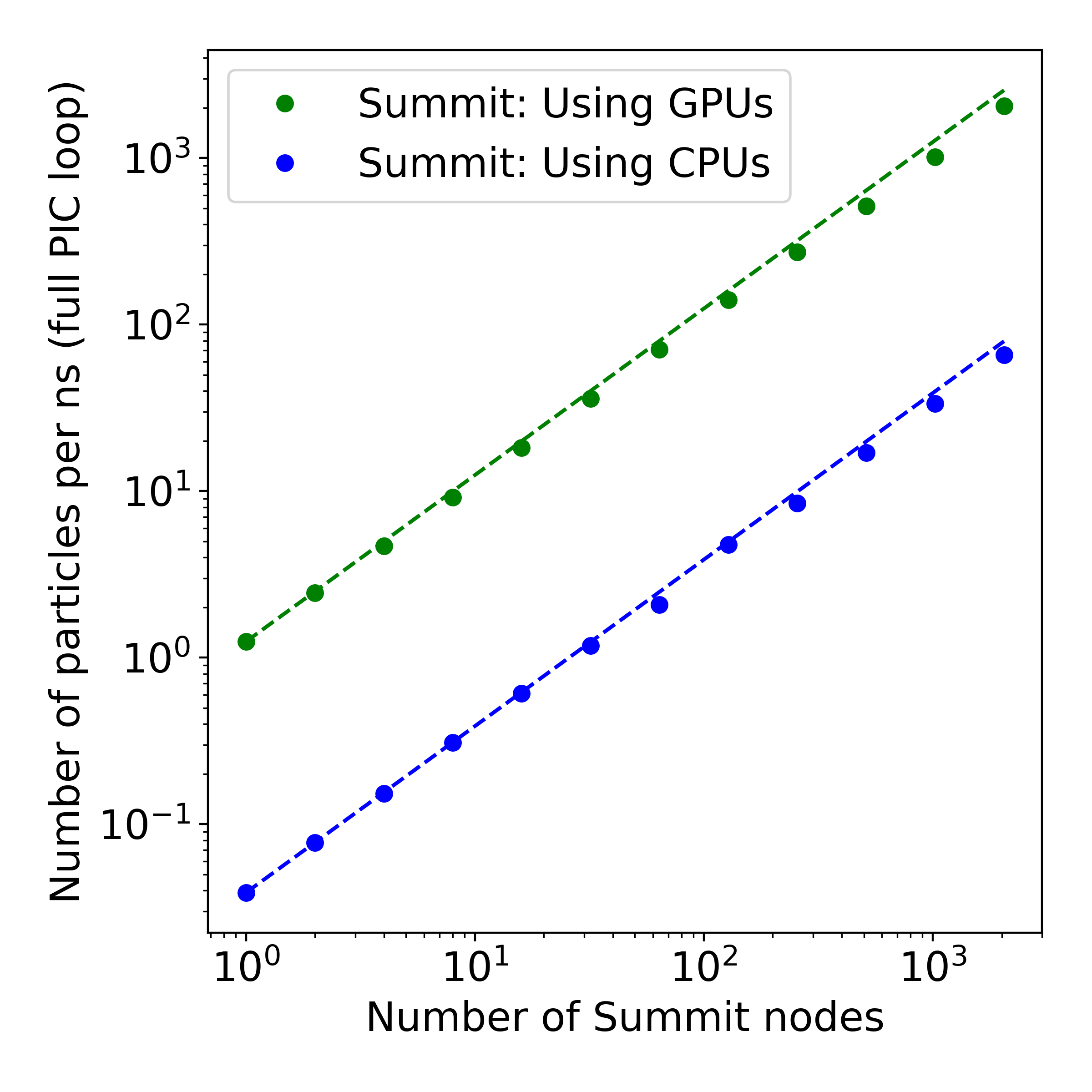}
\end{figure}

The results are shown in Figure \ref{weak_scaling}. We performed the above scaling study twice, once using all six GPUs per Summit node, and again using only the POWER9 CPUs. All CPU and GPU results presented in this section used versions of WarpX \footnote{WarpX Version: 20.10-58-g7a3d26f1cc8d} and AMReX \footnote{AMReX Version: 20.10-47-gf29a0c9d1b8e} from 10/2020, in which all the optimizations discussed in Section \ref{sec:lessons} were present. For both runs, we used 6 MPI tasks per node. For the GPU-accelerated runs, we used one GPU per MPI task, and for the CPU-only case, we used 7 OpenMP threads per task, so that all 42 physical cores on the node were active. Note that, while the POWER9 CPUs on Summit are capable of simultaneous multi-threading (SMT) - running more than 1 hardware thread per physical core - we do not typically see a large benefit to using this feature with WarpX. To confirm this trend for this problem setup, we have taken the 1 node version of the problem above and also run it using 2 (SMT2) and 4 (SMT4) hardware threads per physical core. The SMT2 run was approximately $2.4\%$ faster than without using SMT, while the SMT4 run was $13.9\%$ slower. Thus, while there is a small benefit to using SMT on this problem, using it would not significantly alter our conclusions here. Likewise, we have experimented with different combinations of MPI ranks per node and OpenMP threads per rank other than 6-7 split shown in Figure \ref{weak_scaling}. Using one MPI rank per socket rather than 3 and 21 OpenMP threads per rank gives the same timings to within $0.2\%$, while other combinations, such as 1 MPI and 42 OpenMP threads per node, were slower by a few percentage points. 

Using these results, we can characterize both the weak scaling behavior of the CPU and GPU versions of the WarpX, as well as see the overall speedup obtained on Summit from using the accelerators. In both cases, the code scales well up to 2048 nodes. The weak scaling efficiency, defined as the total time taken for 100 time steps on 1 node divided by the total taken on 2048 nodes, is 81\% for the GPU case and 90\% for the CPU case. The difference in scaling efficiency between the CPU and GPU can be attributed to the fact that, because the local work is significantly faster when using the V100s, communication operations like \texttt{FillBoundary}, which are inherently harder to scale, become relatively more expensive. Additionally, the speedup from the accelerators at all scales tested was a factor of ~30. This speedup refers to the total run time, including time associated with host / device memory traffic and communication, not to isolated compute kernels.

\subsubsection{Strong Scaling study}

We have also conducted a series of strong scaling tests, using a very similar uniform plasma problem setup as before. The only difference is that the box size has been set to $64^3$, to allow for more GPUs / MPI tasks to be used as the problem is strong scaled. There is some overhead associated with doing this, since with smaller boxes, the surface to volume ratio of ghost cells is higher. Other than the box size, the parameters are all the same as before.

We use a series of problem sizes, each scaled up a factor of 2 in terms of the number of cells and particles in the domain. For each one, we conduct a series of five runs, increasing the number of MPI tasks by a factor of 2 each time. Thus, in the fifth run, the run time should have decreased by a factor of 16, assuming perfect strong scaling. By the time we have multiplied the number of MPI ranks by 16, this problem has reached the point where the compute work and the communication work take approximately the same amount of time, so we would not expect the problem to scale further than that. 

\begin{figure}[h]
\caption{\label{fig:strong_scaling} Strong scaling studies for a variety of problem sizes. Each tick type refers to a different problem size. The $x$-axis shows the number of Summit nodes, and the $y$-axis shows scaling efficiency, defined as the time a run should take assuming perfect strong scaling within a problem size \textit{and} perfect weak scaling from the base problem size, divided by the actual run time.}
\centering
\includegraphics[width=0.45\textwidth]{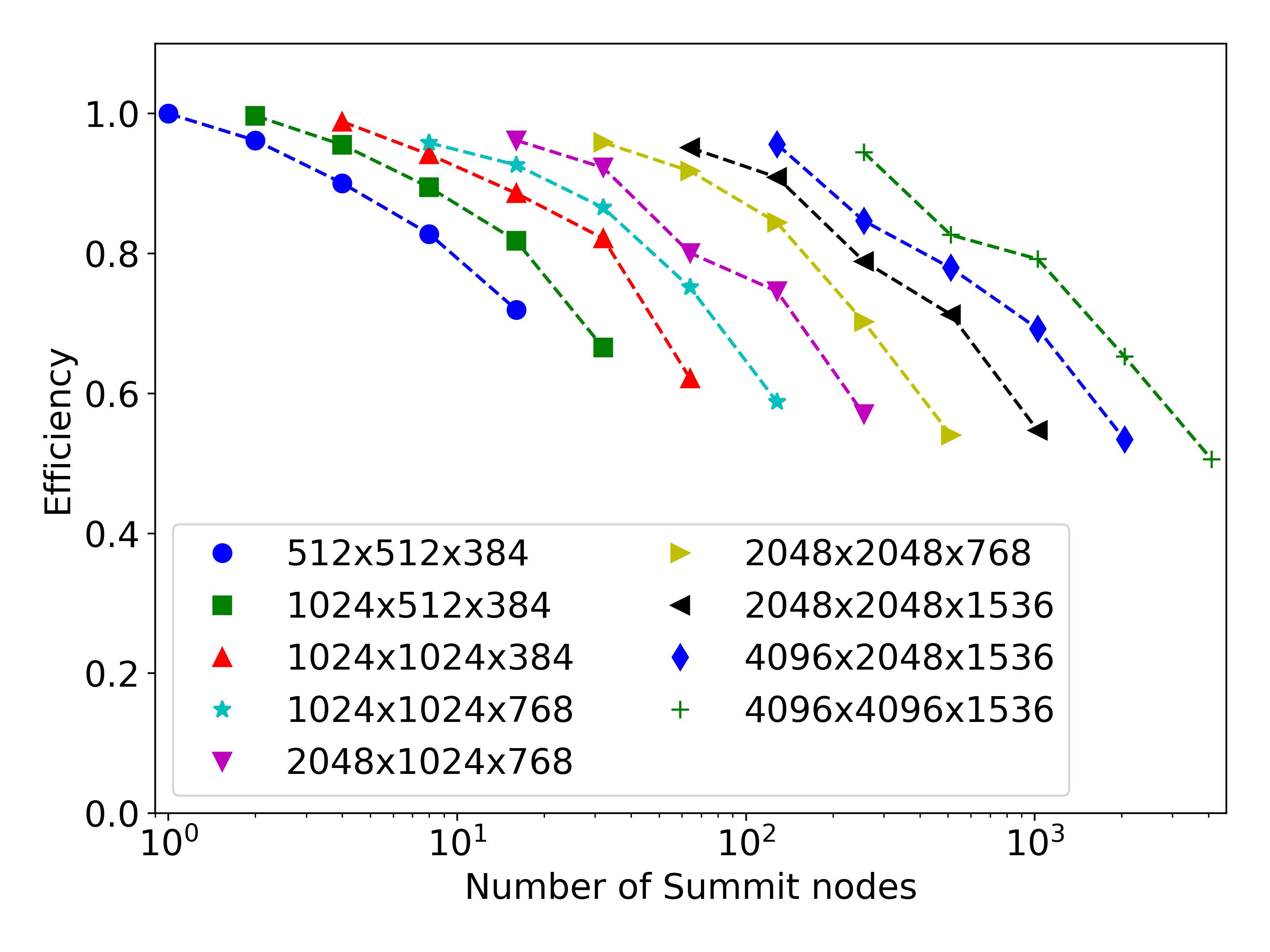}
\end{figure}

The smallest scaling study in this series goes from 1 to 16 nodes, while the largest goes from 256 to 4096, nearly the entire machine. The scaling efficiency, defined as the time a run should take assuming perfect strong scaling within a problem size \textit{and} perfect weak scaling from the base problem size divided by the actual run time, is plotted in Figure \ref{fig:strong_scaling}. The efficiencies after strong scaling by a factor of 16 for each problem size vary from approximately 70\% for the smallest case to approximately 50\% for the largest. 

\subsection{Plasma Acceleration Stage}

The above tests were highly idealized in several ways. First, the workload was perfectly uniform at initial time, and approximately uniform at later times, subject only to random fluctuations in the particle density from cell to cell. Second, the number of particles per cell, 8, is significantly higher than used in some WarpX physics applications. Laser-wake\-field acceleration runs, for example, tend to use about 2 particles per cell on average, which can change the performance profile of the code. Evaluating WarpX on this important science scenario, the following setup was used, designed to mimic the essential features of modelling a single plasma-accelerator stage from WarpX's challenge problem. This is also the benchmark problem used to determine a Figure-of-Merit (FOM) for the ECP Key-Performance Parameters (KPP) assessment. As a KPP-1 project, WarpX needs to show at least a factor of 50 increase in its FOM over the baseline on the eventual Exascale hardware. In this setup, an accelerated particle beam is tracked using the moving window feature in WarpX, in which the simulation domain itself shifts along with the beam at speed c. Additionally, the entire simulation is modeled in a Lorentz-boosted reference frame \cite{Vayprl07}, using a gamma boost of 30. New plasma is continuously injected at the right-hand side of the domain, while particles that leave the domain at the left-hand side are removed from the simulation. The plasma consists of two particles per cell (one electron and one proton), while the accelerated beam is comprised of electrons. Mitigating the numerical Cherenkov instability in the modeling of a relativistically flowing plasma, the Godrey filter \cite{Godfrey2014} is applied to the electromagnetic fields prior to gathering them to particle positions. For the algorithmic options, we have used the Vay particle pusher \cite{vay_pusher}, the Cole-Karkkainen-Cowan FDTD solver \cite{Karkicap06}, and energy-conserving field gathering. We have again used Esirkepov current deposition with 3rd-order interpolation. To minimize the computer time needed to conduct these simulations, we initialize the problem to have the simulation domain entirely filled with plasma, which would normally not be the case when modelling an accelerator stage. 

To gauge the impact of using accelerated nodes on this more realistic problem setup, we have measured the FOM on Summit, defined as 
\begin{equation}
		\text{FOM} = \text{num\_cells} * (\alpha + \beta*\text{ppc}) / \text{avg\_time\_per\_it}
\end{equation}
		
where $\text{num\_cells}$ is the total number of grid points in the simulation, $\alpha$ is 0.1 as heuristic grid update cost, $\beta$ is 0.9 for particle update costs, ppc is the average number of particles per cell, and $\text{avg\_time\_per\_it}$ is the average time per iteration after 1000 steps. We performed this measurement on 4263 Summit nodes, and extrapolated this number to the full machine assuming perfect weak scaling. Our baseline FOM was measured on NERSC's Cori using the original Warp code. The baseline FOM value, measured in March 2019 on 6625 Cori nodes and extrapolated to the 9668 on the full machine, was 2.2e10. The corresponding value on Summit, measured in July 2020, was 2.5e12, over a factor of 100 improvement from the baseline. Additionally, the best CPU-only FOM obtained using the WarpX code was 1.0e11, also measured in March 2019. So there is a substantial (25x) improvement in our FOM measured with WarpX from using the GPUs on Summit, as compared to Cori. 

These values are all summarized in Table \ref{table:FOM}, along with several other data points showing the evolution of WarpX's FOM over time. Of particular interest, the improvement from 9/19 to 1/20 was mostly due to optimizations in the parallel communication routines in AMReX (Section \ref{sec:comm_opt}); from 1/20 to 2/20, the addition of the particle sorting described in Section \ref{particle_sorting}; from 2/20 to 6/20, the reduction in the size of the particle data described in Section \ref{sec:mem_opt}. Finally, the improvement from 6/20 to 7/20 was solely due to being able to run a problem with more cells per node. This illustrates the point made in Section \ref{sec:mem_opt}, about the importance of reducing the memory footprint on Summit. Both by reducing the size of the WarpX particle data and by reducing overhead in AMReX's Arenas, an overall larger problem was able to be run on Summit, resulting in a more efficient use of the machine.

\begingroup
\setlength{\tabcolsep}{5pt}
\begin{table}[width=\linewidth,cols=4,pos=h]
\caption{\label{table:FOM} Progress in the FOM measurement over time. Code: either the original Warp code (baseline) or WarpX. Date: the date when the measurement was taken. Machine: which computer was used to make the measurement. $N_c$/Node: the problem size in number of cells per node. Nodes: how many nodes the measurement was performed on; there are 9668 KNL nodes on Cori and 4608 nodes on Summit. FOM: the figure of merit, extrapolated from the number of nodes the measurement was taken on to the full machine. }\label{tbl1}
\begin{tabular*}{\tblwidth}{@{} LLLLLL@{} }
\toprule
Code & Date & Machine & N$_c$/Node & Nodes & FOM \\
\midrule
Warp  & 3/19 & Cori   & 0.4e7 & 6625 & 2.2e10 \\
WarpX & 3/19 & Cori   & 0.4e7 & 6625 & 1.0e11 \\
WarpX & 6/19 & Summit & 2.9e7 & 32   & 8.6e11 \\
WarpX & 6/19 & Summit & 2.8e7 & 1000 & 7.8e11 \\
WarpX & 9/19 & Summit & 2.3e7 & 2560 & 6.8e11 \\
WarpX & 1/20 & Summit & 2.3e7 & 2560 & 1.0e12 \\
WarpX & 2/20 & Summit & 2.5e7 & 4263 & 1.2e12 \\
WarpX & 6/20 & Summit & 2.0e7 & 4263 & 1.4e12 \\
WarpX & 7/20 & Summit & 2.0e8 & 4263 & 2.5e12 \\
% post paper measurements
%WarpX & 3/21 & Summit & 2.0e8 & 4263 & 2.7e12\\
\bottomrule
\end{tabular*}
\end{table}
\endgroup

\section{Conclusion}

We have summarized the approach taken to porting\linebreak[4] WarpX, which was originally designed for many-core CPU architectures, to take advantage of GPU-accelerated nodes. This approach is largely based on the \verb+amrex::ParallelFor+ set of performance portability functions. We have summarized several key lessons learned from the port, including the importance of managing memory allocation and the code's overall memory footprint, the importance of minimizing the effect of kernel launch latency in MPI communication routines, and the importance of utilizing the cache hierarchy on V100 GPUs. The GPU port of WarpX scales up to nearly all of Summit and currently sees good improvements in its KPP-1 figure of merit on Summit relative to its baseline.

While the measurements of the GPU performance optimizations discussed in this paper were all based on the NVIDIA GPUs on Summit, we expect that most of these (for example, refactoring code to reduce overhead associated with kernel launch latency) will generalize to other accelerator architectures, such as AMD's MI100, as well. Additionally, the optimizations based on reducing memory footprint and exploiting cache hierarchy should transfer to CPU architectures as well. Finally, although the AMReX \texttt{ParallelFor} targets CUDA, HIP, and DPC++ as parallel backends, the optimizations discussed here are not specific to those programming models and should transfer to other approaches such as OpenACC and OpenMP.

\section{Acknowledgements}

This research was supported by the Exascale Computing Project (17-SC-20-SC), a joint project of the U.S. Department of Energy’s Office of Science and National Nuclear Security Administration, responsible for delivering a capable exascale ecosystem, including software, applications, and hardware technology, to support the nation’s exascale computing imperative. This work was performed in part under the auspices of the U.S. Department of Energy by Lawrence Berkeley National Laboratory under Contract\linebreak[4] DE-AC02-05CH11231, SLAC National Accelerator Laboratory under contract AC02-76SF00515, and Lawrence Livermore National Laboratory under Contract\linebreak[4] DE-AC52-07NA27344.

This research used resources of the Oak Ridge Leadership Computing Facility, which is a DOE Office of Science User Facility supported under Contract\linebreak[4] DE-AC05-00OR22725.

WarpX is developed as open source project and available under \url{https://github.com/ECP-WarpX/WarpX}. Presented code versions correspond to the monthly releases of the code between 3/2019 and 10/2020.
The data that support the findings of this study are available under \href{https://doi.org/10.5281/zenodo.4277941}{DOI:10.5281/\linebreak[4]zenodo.4277941}.

\printcredits

\bibliographystyle{cas-model2-names}

\bibliography{warpx}


\end{document}